\documentclass{article}
\usepackage[totalwidth=13.0cm,totalheight=20.0cm]{geometry}
\usepackage{latexsym,amsthm,amsmath,amssymb,url,textcomp}
\usepackage{amsfonts}
\usepackage{graphicx}
\usepackage{subfigure}
\usepackage[USenglish]{babel}
\usepackage{multirow}

\newtheorem{lemma}{Lemma}

\newtheorem{theorem}{Theorem}
\newtheorem{proposition}{Proposition}

{\bfseries}{\itshape}
\newcommand{\dom}{\mbox{$\rightarrow$}}

\begin{document}

\title{Every Ternary Permutation Constraint Satisfaction Problem Parameterized Above Average Has a Kernel with a Quadratic Number of Variables\thanks{Extended abstract of this paper has appeared in the proceedings of the 18th Annual European Symposium on Algorithms, 2010. Part of this research has been supported by the EPSRC, grant EP/E034985/1, the Netherlands Organisation for Scientific Research (NWO), grant 639.033.403, and the Allan Wilson Centre for Molecular Ecology and Evolution.}}

\author{Gregory Gutin\inst{1}, Leo van Iersel\inst{2}, Matthias Mnich\inst{3}, Anders Yeo\inst{1}}

\author{Gregory Gutin${}^1$, Leo van Iersel${}^2$, Matthias Mnich${}^3$, and Anders Yeo${}^1$\\[6pt]
\small ${}^1$  Royal Holloway, University of London\\[-3pt]
\small Egham, Surrey TW20 0EX, UK\\[-3pt]
\small \texttt{\{gutin|anders\}@cs.rhul.ac.uk}\\
\small ${}^2$ University of Canterbury
\small Christchurch, New Zealand\\[-3pt]
\small \texttt{l.j.j.v.iersel@gmail.com}\\[-3pt]
\small ${}^3$ Technische Universiteit Eindhoven
\small Eindhoven, The Netherlands\\[-3pt]
\small \texttt{m.mnich@tue.nl}
}

\maketitle

\begin{abstract}
A ternary Permutation-CSP is specified by a subset $\Pi$ of the symmetric group $\mathcal S_3$.
An instance of such a problem
consists of a set of variables $V$ and a multiset of constraints, which are
ordered triples of distinct variables of $V.$ The objective is
to find a linear ordering $\alpha$ of $V$ that maximizes
the number of triples whose rearrangement (under $\alpha$)
follows a permutation in $\Pi$. We prove that every ternary Permutation-CSP parameterized above average has a kernel with
a quadratic number of variables.
\end{abstract}

\section{Introduction}
\label{sec:introduction}

For maximization problems whose lower bound on the solution value is
a monotonically increasing unbounded function of the instance size, the standard
parameterization by solution value is trivially fixed-parameter tractable.
(Basic notions on parameterized algorithmics used in this paper are given in Section \ref{sec:def}.)
Mahajan and Raman \cite{Mahajan99} were the first to recognize both practical and theoretical importance of parameterizing maximization problems differently:
above tight lower bounds.
They considered {\sc Max Sat} with the tight lower bound $m/2$, where $m$ is the number of clauses, and the problem is to decide whether we can satisfy at least $m/2+k$ clauses, where $k$ is the parameter. Mahajan and Raman proved that this parameterization of {\sc Max Sat} is fixed-parameter tractable by obtaining a kernel with $O(k)$ variables. Despite clear importance of parameterizations above tight lower bounds, until recently only a few sporadic non-trivial results on the topic were obtained \cite{GKLM,GutinRafieySzeiderYeo07,GutinSzeiderYeo08,Mahajan99,HeggernesPaulTelleVillanger07}.

Massive interest in parameterizations above tight lower bounds came with the
paper of Mahajan et al. \cite{MahajanRamanSikdar09}, who stated several questions on fixed-parameter
tractability of maximization problems parameterized above tight lower bounds,
some of which are still open. Several of those questions were answered by
newly-developed methods \cite{AlonEtAl2010,CrowstonEtAl2009,CGJKR,GutinEtAl2009b,GutinEtAl2009a}, using algebraic, probabilistic and harmonic
analysis tools. In particular, a probabilistic approach allowed Gutin et al.
\cite{GutinEtAl2009b} to prove the existence of a quadratic kernel for the parameterized
{\sc Betweenness Above Average} ({\sc Betweenness-AA}) problem, thus, answering an
open question of Benny Chor \cite{Niedermeier2006}.

{\sc Betweenness} is just one representative of a rich family of \emph{ternary Permutation Constraint Satisfaction Problems (CSPs)}.
A ternary Permutation-CSP is specified by a subset $\Pi$ of the symmetric group $\mathcal S_3$.
An instance of such a problem
consists of a set of variables $V$ and a multiset of constraints, which are
ordered triples of distinct variables of $V.$ The objective is
to find a linear ordering $\alpha$ of $V$ that maximizes
the number of triples whose rearrangement (under $\alpha$)
follows a permutation in $\Pi$.
Important special cases are {\sc Betweenness}~\cite{CharikarEtAl2009,Goerdt-a,GutinEtAl2009b,Opatrny1979} and {\sc Circular Ordering}~\cite{GalilMegiddo1977,Goerdt-b}, which find applications in circuit design and computational biology~\cite{CS98,Opatrny1979}, and in qualitative spatial reasoning \cite{IC00}, respectively.

In this paper, we prove that every ternary Permutation-CSP has a
kernel with a quadratic number of variables, when parameterized above
average (AA), which is a tight lower bound.
This result is obtained by first reducing all the problems to just one, {\sc Linear Ordering-AA}, then showing that {\sc Linear Ordering-AA} has a kernel with a quadratic number of variables and constraints and, thus, concluding that there is a bikernel with a quadratic number of variables from each of the problems AA to {\sc Linear Ordering-AA}. Using the last result, we prove that there is a bikernel with a quadratic number of variables from every ternary Permutation-CSP to most ternary Permutation-CSPs. This implies the existence of kernels with a quadratic number of variables  for most ternary Permutation-CSPs. The remaining ternary Permutation-CSPs are proved to be equivalent to {\sc Acyclic Subdigraph-AA} (a \emph{binary Permutation-CSP} defined in Section \ref{sec:facts}) and since {\sc Acyclic Subdigraph-AA}, as shown in \cite{GutinEtAl2009a}, has a kernel with a quadratic number of variables, the remaining ternary Permutation-CSPs have a kernel with a quadratic number of variables.

The most difficult part of this set of arguments is the proof that {\sc Linear Ordering-AA} has a kernel with a quadratic number of variables and constraints. We can show that if we want to prove this in a similar way as for {\sc Betweenness-AA} (that is, eliminate all instances of {\sc Linear Ordering-AA}
whose optimal solution coincides with the lower bound) we need an infinite number of reduction rules, see Section \ref{sec:inf} for details.
So, determining fixed-parameter tractability of {\sc Linear Ordering-AA} turns out to be much harder than that for {\sc Betweenness-AA}.
Fortunately, we found a nontrivial way of reducing {\sc Linear Ordering-AA} to a combination of {\sc Betweenness-AA} and {\sc Acyclic Subdigraph-AA}. Using further probabilistic and deterministic arguments for the mixed problem, we prove that {\sc Linear Ordering-AA} has a kernel with a quadratic number of variables and constraints.

The rest of the paper is organized as follows. In Section \ref{sec:def}, we give some basic notions on parameterized algorithms and complexity. In Section \ref{sec:permCSP}, we define and discuss ternary Permutation-CSPs; we also reduce all nontrivial ternary Permutation-CSPs AA to {\sc Linear Ordering-AA}. Some $\mathsf{NP}$-hardness results of this section are proved in Section \ref{sec:NP}.
In Section \ref{sec:tools}, we describe probabilistic and harmonic analysis tools used in the paper. In Section \ref{sec:facts}, we obtain some results on {\sc Betweenness-AA} and {\sc Acyclic Subdigraph-AA} needed in the following section, where we prove that {\sc Linear Ordering-AA} has a quadratic kernel. In Section \ref{sec:LAAA}, we also prove our main result, Theorem \ref{thm:main}, that every ternary Permutation-CSs parameterized above average has a kernel with a quadratic number of variables. In Section \ref{sec:inf}, we show that {\sc Linear Ordering-AA} has an infinite number of natural reduction rules.
Finally, in Section~\ref{sec:fr} we state and discuss open problems for further research.

\section{Basics on Parameterized Algorithmics} \label{sec:def}

Parameterized complexity theory is a multivariate framework for a refined analysis of hard
($\mathsf{NP}$-hard) problems, which was introduced by Downey and Fellows in a series of ground breaking papers
in the 1990s \cite{DF1,DF2}.
A \emph{parameterized problem} is a subset $L\subseteq \Sigma^* \times
\mathbb{N}$ over a finite alphabet $\Sigma$; $L$ is
\emph{fixed-parameter tractable} if the membership of an instance
$(I,k)$ in $\Sigma^* \times \mathbb{N}$ can be decided in time
$f(k)\cdot |I|^{O(1)}$ where~$f$ is a function of the
{\em parameter} $k$ only~\cite{DowneyFellows99,FlumGrohe06,Niedermeier2006}. (We would like $f(k)$ to grow as slowly as possible.)

Given a pair $L,L'$ of parameterized problems,
a \emph{bikernelization from $L$ to $L'$} is a polynomial-time
algorithm that maps an instance $(x,k)$ to an instance $(x',k')$ (the
\emph{bikernel}) such that (i)~$(x,k)\in L$ if and only if
$(x',k')\in L'$, (ii)~ $k'\leq h(k)$, and (iii)~$|x'|\leq g(k)$ for some
functions $h$ and $g$. The function $g(k)$ is called the {\em size} of the bikernel.
A {\em kernelization} of a parameterized problem
$L$ is simply a bikernelization from $L$ to itself and a bikernel is a {\em kernel} when $L=L'.$

The notion of a bikernelization was introduced by Alon et al. \cite{AlonEtAl2010},  who observed that a decidable parameterized problem $L$ is fixed-parameter
tractable if and only if it admits a
bikernelization to a decidable parameterized problem $L'$. Not every fixed-parameter
tractable problem has a kernel of polynomial size unless $\mathsf{NP}\subseteq \mathsf{coNP/poly}$
\cite{BodlaenderEtAl2009a,BodlaenderEtAl2009,Bodlaender09}; low degree polynomial size kernels are of main interest due to applications.

\section{Permutation CSPs Parameterized Above Average}\label{sec:permCSP}

Let $V$ be a set of $n$ variables.
A \emph{linear ordering of $V$} is a bijection $\alpha: V \rightarrow [n]$, where $[n]=\{1,2,\ldots ,n\}$.
The symmetric group on three elements is $\mathcal S_3=\{(123),(132),(213),(231),(312),(321)\}$.
A \emph{constraint set over $V$} is a multiset $\mathcal C$ of \emph{constraints}, which are permutations of three distinct elements of $V$.
For each subset $\Pi\subseteq \mathcal S_3$ and a linear ordering $\alpha$ of $V$, a constraint $(v_1,v_2,v_3)\in\mathcal C$ is \emph{$\Pi$-satisfied by $\alpha$} if
there is a permutation $\pi\in \Pi$ such that $\alpha(v_{\pi(1)})<\alpha(v_{\pi(2)})<\alpha(v_{\pi(3)})$. If $\Pi$ is fixed, we will simply say that $(v_1,v_2,v_3)\in\mathcal C$ is \emph{satisfied by $\alpha$}.

For each subset $\Pi\subseteq \mathcal S_3$, the problem {\sc $\Pi$-CSP} is to decide whether for a given pair $(V, \mathcal C)$ of variables and constraints there is a linear ordering $\alpha$ of $V$ that $\Pi$-satisfies all constraints in $\mathcal C$.
A complete dichotomy of the {\sc $\Pi$-CSP} problems with respect to their computational complexity was given by Guttmann and Maucher~\cite{GuttmannMaucher2006}.
For that, they reduced $2^{|\mathcal S_3|}=64$ problems by two types of symmetry.
First, two problems differing just by a consistent renaming of the elements of their permutations are of the same complexity.
Second, two problems differing just by reversing their permutations are of the same complexity.
The symmetric reductions leave 13 problems {\sc $\Pi_i$-CSP}, $i=0,1,\hdots,12$, whose time complexity is
polynomial for $\Pi_{11} = \emptyset$ and $\Pi_{12} = \mathcal{S}_3$ and was otherwise established by Guttmann and Maucher~\cite{GuttmannMaucher2006}, see Table~\ref{tab:allproblems}.

\begin{table}
\centering
  \begin{tabular}{llc}
                                                   &                           & ~~Complexity to Sa-\\
    $\Pi\subseteq \mathcal S_3$~~                  & ~~Common Problem Name~~   & ~~isfy All Constraints\\
    \noalign{\smallskip}
    \hline
    \noalign{\smallskip}
    $\Pi_0=\{(123)\}$       & ~~{\sc Linear Ordering}                       & polynomial\\[0.1cm]
    $\Pi_1 =\{(123), (132)\}$       &  ~~                  & polynomial\\[0.1cm]
    $\Pi_{2} =\{(123), (213), (231)\}$       &  ~~                  & polynomial\\[0.1cm]
    $\Pi_{3} =\{(132),  (231), (312), (321)\}$       &  ~~                  & polynomial\\[0.1cm]
    $\Pi_4 = \{(123),(231)\}$                      & ~~                        & $\mathsf{NP}$-complete\\[0.1cm]
    $\Pi_5 = \{(123),(321)\}$                      & ~~{\sc Betweenness}       & $\mathsf{NP}$-complete\\[0.1cm]
    $\Pi_6 = \{(123),(132),(231)\}$                & ~~                        & $\mathsf{NP}$-complete\\[0.1cm]
    $\Pi_7 = \{(123),(231),(312)\}$                & ~~{\sc Circular Ordering} & $\mathsf{NP}$-complete\\[0.1cm]
    $\Pi_8 = \mathcal S_3\setminus\{(123),(231)\}$ & ~~                        & $\mathsf{NP}$-complete\\[0.1cm]
    $\Pi_9 = \mathcal S_3\setminus\{(123),(321)\}$ & ~~{\sc Non-Betweenness}   & $\mathsf{NP}$-complete\\[0.1cm]
    $\Pi_{10} = \mathcal S_3\setminus\{(123)\}$       & ~~                        & $\mathsf{NP}$-complete\\[0.1cm]
  \end{tabular}
  \vspace{0.2cm}
  \caption{Ternary Permutation-CSPs (after symmetry considerations)}
\label{tab:allproblems}
\end{table}

The maximization version of {\sc $\Pi_i$-CSP} is the problem {\sc Max-$\Pi_i$-CSP} of finding a linear ordering $\alpha$ of $V$ that $\Pi_i$-satisfies a maximum number of constraints in $\mathcal C$.
Clearly, for $i = 4,\hdots,10$ the problem {\sc Max-$\Pi_i$-CSP} is $\mathsf{NP}$-hard.
In Section \ref{sec:NP} we prove that  {\sc Max-$\Pi_i$-CSP} is $\mathsf{NP}$-hard also for $i=0,1,2,3$.

Now observe that given a variable set $V$ and a constraint multiset $\mathcal C$ over $V$, for a random linear ordering $\alpha$ of $V$, the probability of a constraint in $\mathcal C$ being $\Pi$-satisfied by $\alpha$ equals $\frac{|\Pi|}{6}$.
Hence, the expected number of satisfied constraints from $\mathcal C$ is $\frac{|\Pi|}{6}|\mathcal C|$, and thus there is a linear ordering $\alpha$ of $V$ satisfying at least $\frac{|\Pi|}{6}|\mathcal C|$ constraints (and this bound is tight). A derandomization argument leads to $\frac{|\Pi_i|}{6}$-approximation algorithms for the problems {\sc Max-$\Pi_i$-CSP} \cite{CharikarEtAl2009}. No better constant factor approximation is possible assuming the Unique Games Conjecture~\cite{CharikarEtAl2009}.

We study the parameterization of {\sc Max-$\Pi_i$-CSP} above tight lower bound:

\medskip
\noindent \begin{tabular}{lp{0.85\textwidth}}
\multicolumn{2}{l}{{\sc $\Pi$-Above Average ($\Pi$-AA)}}   \\
\textit{Input:}     & A finite set $V$ of variables, a multiset $\mathcal C$ of ordered trip\-les  of distinct variables from $V$ and an integer $k\geq 0$.                   \\
\textit{Parameter:} & $k$.\\
\textit{Question:}  & Is there a linear ordering $\alpha$ of $V$ such that at least $\frac{|\Pi|}{6}|\mathcal C| + k$ constraints of $\mathcal C$ are $\Pi$-satisfied by $\alpha$?\\
\end{tabular}
\medskip

For example, choose $\Pi = \{(123),(321)\}$ for {\sc Betweenness-AA}. We will call
$\Pi_0$-AA the {\sc Linear Ordering-AA} problem.

Let~$\Pi$ be a subset of $\mathcal S_3$.
Clearly, if $\Pi$ is the empty set or equal to $\mathcal S_3$ then the corresponding problem $\Pi$-AA can be solved in polynomial time.
The following simple result allows us to study the {\sc $\Pi$-AA} problems using {\sc $\Pi_0$-AA}.

\begin{proposition}
\label{thm:allreducetoone}
  Let $\Pi$ be a subset of $\mathcal S_3$ such that $\Pi\notin\{\emptyset,\mathcal S_3\}$.
  There is a polynomial time transformation $f$ from $\Pi$-AA to $\Pi_0$-AA such that an instance $(V,{\mathcal C},k)$ of $\Pi$-AA is a ``yes''-instance if and only if $(V,{\mathcal C}_0,k)=f(V,{\mathcal C},k)$ is a ``yes''-instance of $\Pi_0$-AA.
\end{proposition}
\begin{proof}
  From an instance $(V,\mathcal C,k)$ of~$\Pi$-AA, construct an instance $(V,\mathcal C_0,k)$ of~$\Pi_0$-AA as follows.
  For each triple $(v_1,v_2,v_3)\in\mathcal C$, add $|\Pi|$ triples $(v_{\pi(1)},v_{\pi(2)},v_{\pi(3)})$, $\pi\in \Pi$, to~$\mathcal C_0$.

  Observe that a triple $(v_1,v_2,v_3)\in\mathcal C$ is $\Pi$-satisfied if and only if exactly one of the triples $(v_{\pi(1)},v_{\pi(2)},v_{\pi(3)})$, $\pi\in \Pi$, is $\Pi_0$-satisfied.
  Thus, $\frac{|\Pi|}{6}|\mathcal C| + k$ constraints from $\mathcal C$ are $\Pi$-satisfied if and only if the same number of constraints from $\mathcal C_0$ are $\Pi_0$-satisfied.
  It remains to observe that $\frac{|\Pi|}{6}|\mathcal C| + k=\frac{1}{6}|\mathcal C_0| + k$ as $|\mathcal C_0|=|\Pi|\cdot |\mathcal C|$.
\end{proof}

For a variable set $V$, a constraint multiset $\mathcal C$ over $V$ and a linear ordering $\alpha$ of $V$, the \emph{$\alpha$-deviation of $(V, \mathcal C)$} is the number $\mathsf{dev}(V,{\mathcal C},\alpha)$ of constraints of $\mathcal C$ that are $\Pi$-satisfied by $\alpha$ minus $\frac{|\Pi|}{6}|\mathcal C|$.
The \emph{maximum deviation of $(V,\mathcal C)$},
denoted $\mathsf{dev}(V,\mathcal C)$, is the maximum of
$\mathsf{dev}(V,{\mathcal C},\alpha)$ over all linear orderings
$\alpha$ of $V$.
Now the problem $\Pi$-AA can be reformulated as the problem of deciding whether $\mathsf{dev}(V,{\mathcal C})\ge k$.

\section{$\mathsf{NP}$-hardness of {\sc Max-$\Pi_i$-CSP} for $i=0,1,2,3$}\label{sec:NP}

The problem {\sc Acyclic Subdigraph} is, given a directed multigraph $D$ and an integer $k > 0$, to decide whether $D$ contains an acyclic subdigraph with at least $k$ arcs.
{\sc Acyclic Subdigraph} can be reformulated as a problem of verifying whether $V$ has a linear ordering $\alpha$ in which at least $k$ arcs are {\em satisfied}, i.e., for each such arc $(u,v)$ we have $\alpha(u)<\alpha(v).$
It is well-known that {\sc Acyclic Subdigraph} is $\mathsf{NP}$-complete.

\begin{theorem}
\label{thm:allNPhard}
For $i = 0,1,2,3$, problem {\sc Max-$\Pi_i$-CSP} from Table \ref{tab:allproblems} is $\mathsf{NP}$-hard.
\end{theorem}
\begin{proof}
We will consider the four cases one by one.
\begin{description}
  \item[$i=0$:]
    Proposition \ref{thm:allreducetoone} implies, in particular, that {\sc Max-Betweenness} can be reduced to {\sc Max-$\Pi_0$-CSP}. Thus, {\sc Max-$\Pi_0$-CSP} is $\mathsf{NP}$-hard.

\item[$i=1:$]
    Denote constraints of {\sc Max-$\Pi_1$-CSP} by $(u<\min\{v,w\})$.
    Such a constraint is $\Pi_1$-satisfied by a linear ordering $\alpha$ of $\{u,v,w\}$ if and only if $\alpha(u)<\min\{\alpha(v),\alpha(w)\}$.
   From an instance $(D=(U,A),k)$ of {\sc Acyclic Subdigraph}, we construct an instance $(V,{\mathcal C},k)$ of (a decision version of) {\sc Max-$\Pi_1$-CSP} by setting $V = U \cup \{z\}$ and,
  for each arc $(u,v)\in A$, adding $(u < \min\{v,z\})$ to $\mathcal C$.
  Observe that, without loss of generality, an optimal linear ordering of $(V,\mathcal C,k)$ has $z$ at the end as if it does not then moving
$z$ to the end does not falsify any constraints. Therefore $(u,v)$ is satisfied in $D$ if and only
if $(u < \min\{v,z\})$ is $\Pi_1$-satisfied in $(V,\mathcal C,k)$.
   \item[$i=2:$]
    Denote constraints of {\sc Max-$\Pi_2$-CSP} by $(u,v<w)$.
    Such a constraint is $\Pi_2$-satisfied by a linear ordering $\alpha$ of $\{u,v,w\}$ if and only if $\alpha(v) < \alpha(w)$.
    From an instance $(D = (U,A),k)$ of {\sc Acyclic Subdigraph}, we construct an instance $(V,{\mathcal C},k)$ of (a decision version of) {\sc Max-$\Pi_2$-CSP} by setting $V=U\cup \{z\}$ and, for each arc $(v,w)\in A$, adding constraint $(z,v<w)$ to the constraint set $\mathcal C$.
    Observe that $D$ has a set of $k$ arcs that form an acyclic subdigraph if and only if there are $k$ constraints in $\mathcal C$ that can be $\Pi_2$-satisfied by a linear ordering of $V$.
    Thus, we have reduced {\sc Acyclic Subdigraph} to {\sc Max-$\Pi_2$-CSP}, implying that {\sc Max-$\Pi_2$-CSP} is $\mathsf{NP}$-hard.
  \item[$i=3:$] Let us denote a constraint in MAX-$\Pi_3$-CSP by $(\max\{u,v\} \not< w)$.
This constraint is $\Pi_3$-satisfied by a linear ordering $\alpha$ if and only if $w$ is not the last element among $u,v,w$ in $\alpha$.
  Now consider an instance $(V,\mathcal C_1,k)$ of MAX-$\Pi_1$-CSP, which we have shown to be $\mathsf{NP}$-hard.
For each constraint $(u < \min\{v,w\})$ in $\mathcal C_1$ add $(\max\{u,v\} \not< w)$ and $(\max\{u,w\} \not< v)$ to $\mathcal C_3$. Now
we will show that $(V,\mathcal C_1,k)$ is a ``yes"-instance if and only if $(V,\mathcal C_3, |\mathcal C_1|+k)$ is a ``yes"-instance of MAX-$\Pi_3$-CSP.
Let $\alpha$ be any linear ordering
of $V$ and let $\alpha'$ be the reverse ordering. Note that $(u < \min\{v,w\})$ is $\Pi_1$-satisfied by $\alpha$ if and
only if both $(\max\{u,v\} \not< w)$ and $(\max\{u,w\} \not< v)$ are $\Pi_3$-satisfied by $\alpha'$. Furthermore one of
$(\max\{u,v\} \not< w)$ and $(\max\{u,w\} \not< v)$ is always $\Pi_3$-satisfied. Therefore, at least $k$ constraints of $\mathcal C_1$ are $\Pi_1$-satisfied by $\alpha$
 if and only if at least $2k+(|\mathcal C_1|-k)$ constraints of $\mathcal C_3$ are $\Pi_3$-satisfied by $\alpha'$. So, we have reduced MAX-$\Pi_1$-CSP to MAX-$\Pi_3$-CSP,
and we are done.
\end{description}
\end{proof}

\section{Probabilistic and Harmonic Analysis Tools}
\label{sec:tools}

We build on the probabilistic \emph{Strictly Above Expectation} method by Gutin et al.~\cite{GutinEtAl2009a} to prove non-trivial lower bounds on the minimum fraction of satisfiable constraints in instances belonging to a restricted subclass.
For such an instance with parameter~$k$, we introduce a random variable $X$ such that the instance is a ``yes''-instance if and only if $X$ takes with positive probability a value greater than or equal to~$k$.
If $X$ happens to be a symmetric random variable with finite second moment then $\mathbb P(X \geq \sqrt{\mathbb{E}[X^2]}) > 0$; it hence suffices to prove $\mathbb E[X^2] = h(k)$ for some monotonically increasing unbounded function $h$.
(Here, $\mathbb P(\cdot)$ and $\mathbb E[\cdot]$ denote probability and expectation, respectively.)
If $X$ is not symmetric then the following lemma can be used instead.
\begin{lemma}[Alon et al.~\cite{AlonEtAl2010}]
\label{lem:moments}
  Let $X$ be a real random variable and suppose that its first, second and forth moments satisfy $\mathbb{E}[X] = 0$, $\mathbb{E}[X^2] = \sigma^2 > 0$ and $\mathbb{E}[X^4] \leq c \sigma^4$, respectively, for some constant $c$.
  Then $\mathbb P(X > \frac{\sigma}{2 \sqrt c}) > 0$.
\end{lemma}

\noindent
We combine this result with the following result from harmonic analysis.

\begin{lemma}[Hypercontractive Inequality~\cite{Bonami1970,Gross1975}]
\label{lem:polynomial}
  Let $f = f(x_1,\ldots,x_n)$ be a polynomial of degree $r$ in $n$ variables $x_1,\ldots,x_n$ with domain $\{-1,1\}$.
  Define a random variable $X$ by choosing a vector $(\epsilon_1,\ldots,\epsilon_n)\in \{-1,1\}^n$ uniformly at random and setting $X = f(\epsilon_1,\ldots,\epsilon_n)$.
  Then $\mathbb E[X^4]\leq 9^r\mathbb E[X^2]^2$.
\end{lemma}

\section{Betweenness and Acyclic Subdigraph Problems}\label{sec:facts}

Let $u,v,w$ be variables.
We denote a betweenness constraint ``$v$ is between $u$ and $w$'' by $(v,\{u,w\})$, and call a 3-set $S$ of betweenness constraints over $\{u,v,w\}$ \emph{complete} if $S = \{(u,\{v,w\}),(v,\{u,w\}),(w,\{u,v\})\}$. Since every linear ordering of $\{u,v,w\}$ satisfies exactly one constraint in $S$, we obtain the following reduction.
\begin{lemma}\label{thm:reduce_3sets}
  Let $(V,{\mathcal B})$ be an instance of {\sc Betweenness} and let $\alpha$ be a linear ordering of $V$.
  Let $\mathcal B'$ be the set of constraints obtained from ${\mathcal B}$ by deleting all complete subsets.
  Then $\mathsf{dev}(V,{\mathcal B},\alpha) = \mathsf{dev}(V,{\mathcal B}',\alpha)$.
\end{lemma}
An instance of {\sc Betweenness} without complete subsets of constraints is called~\emph{reduced}.

Let $(V,{\mathcal B})$ be an instance of {\sc Betweenness}, with $\mathcal B=\{B_1,\ldots ,B_m\}$, and let $\phi$ be a fixed function from $V$ to $\{0,1,2,3\}$.
A linear ordering $\alpha$ of $V$ is called \emph{$\phi$-compatible} if for each pair $u,v \in V$ with $\alpha (u) < \alpha (v)$ it holds $\phi(u) \leq \phi(v)$.
For a random $\phi$-compatible linear ordering $\pi$ of $V$, define a binary random variable $y_p$ that takes value one if and only if $B_p\in {\mathcal B}$ is satisfied by $\pi$ (if $B_p$
is falsified by $\pi$, then $y_p=0$).
Let $Y_p=\mathbb{E}[y_p] -1/3$ for each $p\in [m]$, and let $Y=\sum_{p=1}^mY_p.$

Now let $\phi$ be a random function from $V$ to $\{0,1,2,3\}$.
Then $Y,Y_1,\ldots ,Y_m$ are random variables.
For a constraint $B_p=(v,\{u,w\})$, the distribution of $Y_p$ as it is given in Table \ref{tab:Yp_dist} implies that $\mathbb{E}[Y_p]=0$.
Thus, by linearity of expectation, $\mathbb{E}[Y]=0.$
\begin{table}
  \centering
  \begin{tabular}{cccc}
    \hline
    \noalign{\smallskip}
    $|\{\phi(u),\phi(v), \phi(w)\}|$ & Relation & Value of $Y_p$ & Prob.\\[0.1cm]
    \noalign{\smallskip}
    \hline
    1 & $\phi(u) = \phi(v) = \phi(w)$ & 0 & 1/16 \\[0.1cm]
    2 & $\phi(v)\neq \phi(u)=\phi(w)$ & $-1/3$ & 3/16\\[0.1cm]
    2 & $\phi(v)\in \{\phi(u), \phi(w)\}$ & $1/6$ & 6/16 \\[0.1cm]
    3 & $\phi(v)$ is between $\phi(u)$ and $\phi(w)$ & $2/3$ & 2/16\\[0.1cm]
    3 & $\phi(v)$ is not between $\phi(u)$ and $\phi(w)$ & $-1/3$ & 4/16\\[0.1cm]
    \noalign{\smallskip}
    \hline
  \end{tabular}
  \vspace{0.2cm}
  \caption{Distribution of $Y_p$ for constraint $B_p=(v,\{u,w\})$.}
\label{tab:Yp_dist}
\end{table}

The following lemma was proved by Gutin et al.~\cite{GutinEtAl2009b} for {\sc Betweenness} in which $\mathcal B$ is a set, not a multiset, but a simple modification of its proof gives us the following:

\begin{lemma}
\label{thm:Ysecondmoment}
  For a reduced instance $(V,{\mathcal B})$ of {\sc Betweenness}, $\mathbb{E}[Y^2] \geq \frac{11}{768}m$.
\end{lemma}
\begin{proof}
Observe that $\mathbb{E}[Y^2] =    \sum_{l=1}^m \mathbb{E}[Y_l^2] + \sum_{1\leq l\neq l' \leq m}\mathbb{E}[Y_lY_{l'}].$ Using Table \ref{tab:Yp_dist}, it is easy to see that
$\sum_{l=1}^m \mathbb{E}[Y_l^2]=\frac{88}{768}m.$

Let $U = \{(l,l'):\ B_l,B_{l'}\in\mathcal B, l \not= l'\}$ be the set of all ordered index pairs corresponding to distinct constraints in $\mathcal B$. Let $U^{*}=\{(l,l')\in U:\ vars(B_l) = vars(B_{l'}), B_l\neq B_{l'}\}$ and $U^{**}=\{(l,l')\in U:\ B_l = B_{l'}\}$, where $vars(B_l)$ and $vars(B_{l'})$ are the sets of variables of $B_l$ and $B_{l'},$ respectively. Taking into consideration that $|U^{*}|\le m$ and $|U^{**}|\ge 0,$ similarly to \cite{GutinEtAl2009b}, we obtain that $$\sum_{(l,l')\in U}\mathbb{E}[Y_lY_{l'}]\ge - \frac{66}{768}m - \frac{11}{768}|U^{*}|  + \frac{22}{768}|U^{**}|\ge - \frac{66}{768}m - \frac{11}{768}m=-\frac{77}{768}m.$$ Combining this with $\sum_{l=1}^m \mathbb{E}[Y_l^2]=\frac{88}{768}m$, we get $\mathbb{E}[Y^2] \geq \frac{11}{768}m$.
\end{proof}

Recall that in the {\sc Acyclic Subdigraph} problem we are given a directed multigraph $D=(U,A)$, with parallel arcs allowed, and ask for a linear ordering $\pi$ of $V$ which maximizes the number of satisfied arcs, where an arc $(u,v)\in A$ is \emph{satisfied by $\pi$} if $\pi(u) < \pi(v)$.
If $\pi$ is a uniformly-at-random linear ordering of $V$ then the probability of an arc of $D$ being satisfied is $1/2$.
Thus, there is a linear ordering $\pi$ of $V$ in which the number of satisfied arcs is at least $|A|/2$.
We therefore define, for a digraph $D = (U,A)$ and a linear ordering $\pi$ of $U$, the \emph{$\pi$-deviation} of $D$ as the number of arcs satisfied by $\pi$ minus $|A|/2$, and denote it by $\mathsf{dev}(V,A,\pi)$. In the {\sc Acyclic Subdigraph-AA} problem we are given a directed multigraph $D=(U,A)$ and asked to decide whether there is a linear ordering $\pi$ of $U$ with \emph{$\pi$-deviation} at least $k$, where $k$ is a parameter.

As every linear ordering of $U$ satisfies exactly one of two mutually opposite arcs $(u,v)$ and $(v,u)$, we obtain the following reduction.
\begin{lemma}
\label{thm:reduce_digraph_ordering}
  Let $D = (U,A)$ be a directed multigraph and let $\pi$ be a linear ordering of $V$.
  Let $A'$ be the set of arcs obtained from $A$ by deleting all pairs of mutually opposite arcs.
  Then $\mathsf{dev}(V,A,\pi) = \mathsf{dev}(V,A',\pi)$.
\end{lemma}
A directed multigraph without mutually opposite arcs is called \emph{reduced}.

Let $D = (U,A)$ be a directed multigraph with $A = \{a_1,\hdots, a_m\}$ as multiset of arcs, and let $\phi$ be a fixed function from $U$ to $\{0,1,2,3\}$.
For a random $\phi$-compatible linear ordering $\pi$ of $U$, define a binary random variable $x_p$ that takes value one if and only if $a_p$ is satisfied by $\pi$.
Let $X_p=\mathbb{E}[x_p] -1/2$ for each $p\in [m]$ and let $X=\sum_{p=1}^mX_p$.

Now let $\phi$ be a random function from $U$ to $\{0,1,2,3\}$. Then $X,X_1,\ldots ,X_m$ are random variables.  For an arc $(u,v)$, the distribution of $X_p$ as it is given in Table~\ref{tab:Xp_dist} implies that $\mathbb{E}[X_p]=0$.
Thus, by linearity of expectation, $\mathbb{E}[X]=0.$
\begin{table}
  \centering
  \begin{tabular}{crc}
    \hline
    \noalign{\smallskip}
    Relation between $\phi(u)$ and $\phi(v)$ & Value of $X_p$ & Prob.\\
    \noalign{\smallskip}
    \hline
    \noalign{\smallskip}
    $\phi(u) = \phi(v)$ & 0 & 1/4 \\
    $\phi(u)< \phi(v)$& $1/2$ & 3/8\\
    $\phi(u)>\phi(v)$ & $-1/2$ & 3/8 \\
    \noalign{\smallskip}
    \hline
  \end{tabular}
  \vspace{0.2cm}
  \caption{Distribution of $X_p$ for an arc $(u,v)$.}
\label{tab:Xp_dist}
\end{table}

We have the following analogue of Lemma~\ref{thm:Ysecondmoment}.
\begin{lemma}
\label{thm:Xsecondmoment}
For reduced directed multigraphs $D$ it holds that $\mathbb{E}[X^2] \geq \frac{1}{32}m$.
\end{lemma}
\begin{proof}
  We write $\mathbb{E}[X^2]$ as the sum
  \begin{equation}\label{eq:E21}
    \mathbb{E}[X^2] = \sum_{p=1}^m \mathbb{E}[X_p^2] + \sum_{1\leq p\neq p'\leq m} \mathbb{E}[X_pX_{p'}].
  \end{equation}
  From Table~\ref{tab:Xp_dist} it follows that $\mathbb{E}[X_p^2]= \frac{3}{16}$, and hence it remains to bound the second sum in \eqref{eq:E21}.
  Consider any ordered pair $(a_p,a_{p'})$ of distinct arcs in $D$.
  If $a_p$ and $a_{p'}$ are vertex-disjoint, then clearly $\mathbb{E}[X_pX_{p'}]=0$.
  If $a_p$ and $a_{p'}$ have vertices in common, we define
  \begin{align*}
    S_1(u)   &    = \{(p,p')~|~a_p=(u,x), a_{p'}=(u,y), x,y\in V\}\\
             & \cup \{(p,p')~|~a_p=(x,u), a_{p'}=(y,u), x,y\in V\}\\
    S_2(u)   &    = \{(p,p')~|~a_p=(u,x), a_{p'}=(y,u), x,y\in V\}\\
             & \cup \{(p,p')~|~a_p=(x,u), a_{p'}=(u,y), x,y\in V\}\\
    S_3(u,v) & = \{(p,p')~|~a_p=(u,v), a_{p'}=(u,v)\} \enspace .
  \end{align*}
  By setting $l(u) = |\{ a\in A : a=(u,y), y\in V \}|$ and $r(u) = |\{ a\in A : a=(x,u), x\in V \}|$ it follows that
  \begin{align*}
    |S_1(u)| & = l(u)(l(u)-1) + r(u)(r(u)-1),\\
    |S_2(u)| & = 2l(u)r(u).
  \end{align*}

  Consider a pair $(p,p')\in S_1(u)$, with say $a_p=(u,x),a_{p'}=(u,y)$.
  It is easy to calculate that out of the 64 functions $\phi:\{u,x,y\}\rightarrow\{0,1,2,3\}$, there are 14 functions in which $\phi(u)<\phi(x)$ and $\phi(u)<\phi(y)$.
  Symmetrically, there are 14 functions $\phi$ in which $\phi(u)>\phi(x)$ and $\phi(u)>\phi(y)$.
  In both cases, $X_pX_{p'}=\frac{1}{4}$, by Table~\ref{tab:Xp_dist}.
  Similarly, there are 4 functions $\phi$ in which $\phi(u)<\phi(x)$ and $\phi(u)>\phi(y)$, and 4 functions $\phi$ in which $\phi(u)>\phi(x)$ and $\phi(u)<\phi(y)$; in both cases $X_pX_{p'}=-\frac{1}{4}$.
  For all other functions $\phi$ we have that $X_pX_{p'}=0$, and thus it follows that $\mathbb{E}[X_pX_{p'}]=\frac{5}{64}$ for each pair of arcs $(a_p,a_{p'})$ in $S_1(u)$.

  Similarly, for each pair $(p,p')\in S_2(u)$ it holds that $\mathbb{E}[X_pX_{p'}]=-\frac{5}{64}$, and for each pair $(p,p')\in S_3(u,v)$ it holds that $\mathbb{E}[X_pX_{p'}]=\mathbb{E}[X_p^2]= \frac{3}{16}$.

  Hence,
  \begin{equation*}
  \sum_{1\leq p\neq p'\leq m} \mathbb{E}[X_pX_{p'}] = \sum_{u\in V}\frac{5}{64} |S_1(u)| - \frac{5}{64} |S_2(u)| + \sum_{u,v\in V} w'|S_3(u,v)|,
  \end{equation*}
  with $\frac{5}{64} + \frac{5}{64} + w' = \frac{3}{16}$, because $S_3(u,v)=S_1(u)\cap S_1(v)$.
  Thus, $w'=\frac{1}{32}$, and we obtain
  \begin{align*}
          & \sum_{1\leq p\neq p'\leq m} \mathbb{E}[X_pX_{p'}]\\
        = & \frac{5}{64} \sum_{u\in V} l(u)(l(u)-1) + r(u)(r(u)-1) - 2l(u)r(u) + \sum_{u,v\in V} \frac{1}{32}|S_3(u,v)|\\
        = & \frac{5}{64} \sum_{u\in V} (l(u)-r(u))^2 - l(u) - r(u) + \sum_{u,v\in V} \frac{1}{32}|S_3(u,v)|\\
     \geq & - \frac{5}{64} \sum_{u\in V} l(u) + r(u) = - \frac{10}{64} m,
  \end{align*}
  because each arc contributes exactly one to $\sum_{u\in V} l(u)$ and one to $\sum_{u\in V} r(u)$.
  We conclude that $\mathbb{E}[X^2]\geq \frac{3}{16}m - \frac{10}{64} m= \frac{1}{32}m$.
\end{proof}

The following theorem was proved in \cite{GutinEtAl2009a}.

\begin{theorem}\label{thm:ASDkernel}
{\sc Acyclic Subdigraph-AA} has a kernel with a quadratic number of vertices and arcs.
\end{theorem}

\section{Kernels for $\Pi$-AA Problems}\label{sec:LAAA}
We start from the following key construction of this paper.
With an instance $(V,{\mathcal C})$ of {\sc Linear Ordering}, we associate an instance $(V,\mathcal B)$ of {\sc Betweenness} and two instances $(V,A')$ and $(V,A'')$ of {\sc Acyclic Subdigraph} as follows: If $C_p=(u,v,w)\in {\mathcal C}$, then $B_p=(v,\{u,w\})\in \mathcal B$, $a'_p=(u,v)\in A'$, and $a''_p=(v,w)\in A''$.

\begin{lemma}\label{lem:dev}
  Let $(V,C,k)$ be an instance of {\sc Linear Ordering-AA} and let $\alpha$ be a linear ordering of $V$.
  Then
  \begin{equation*}
    \mathsf{dev}(V,{\mathcal C},\alpha)
      = \frac{1}{2}\left[   \mathsf{dev}(V,A',\alpha)
                          + \mathsf{dev}(V,A'',\alpha)
                          + \mathsf{dev}(V,{\mathcal B},\alpha)
                  \right].
  \end{equation*}
\end{lemma}
\begin{proof}
  For each constraint $C_p=(u,v,w)\in \mathcal C$, define a binary variable $\hat{x}'_p$ that takes value one if and only if $a'_p$ is satisfied by $\alpha$.
  Similarly, define binary variables $\hat{x}''_p$ for arc $a''_p$, $\hat{y}_p$ for constraint $B_p$ and $\hat{z}_p$ for constraint $C_p$.
  To show the lemma it suffices to prove that for each constraint $C_p\in \mathcal C$ and every linear ordering $\pi$ of $\{x,y,z\}$ it holds that
  \begin{equation*}
    \mathsf{dev}(V,\{C_p\},\pi)
      = \frac{1}{2}\left[   \mathsf{dev}(V,\{a'_p\},\pi)
                          + \mathsf{dev}(V,\{a''_p\},\pi)
                          + \mathsf{dev}(V,\{B_p\},\pi)
                   \right],
  \end{equation*}
  where $\mathsf{dev}(V,\{C_p\},\pi)=\hat{z}_p-1/6$, $\mathsf{dev}(V,\{a'_p\},\pi)=\hat{x}'_p-1/2$, $\mathsf{dev}(V,\{a''_p\},\pi)=\hat{x}''_p-1/2$ and $\mathsf{dev}(V,\{B_p\},\pi)=\hat{y}_p-1/3$.
  Thus, it suffices to prove that $\hat{z}_p=(\hat{x}'_p + \hat{x}''_p + \hat{y}_p -1)/2$.
  But this expression holds, as can be seen from Table~\ref{tab:xyzt}: if $C_p$ is satisfied by $\pi$ then all three constraints $a'_p,a''_p,B_p$ are satisfied by $\pi$, whereas if $C_p$ is not satisfied by $\pi$ then exactly one of the three constraints $a'_p,a''_p,B_p$ is satisfied by $\pi$.
  \begin{table}
    \centering
    \begin{tabular}{lll}
      \hline
      \noalign{\smallskip}
      linear ordering $\pi$         &\qquad& constraints\\
      of $\{u,v,w\}$              && satisfied by $\pi$\\
      \hline
      \noalign{\smallskip}
      $uvw$ && $(u,v),(v,w),(v,\{u,w\})$\\[0.1cm]
      $uwv$ && $(u,v)$\\[0.1cm]
      $wuv$ && $(u,v)$\\[0.1cm]
      $vuw$ && $(v,w)$\\[0.1cm]
      $vwu$ && $(v,w)$\\[0.1cm]
      $wvu$ && $(v,\{u,w\})$\\
      \noalign{\smallskip}
      \hline
    \end{tabular}
    \vspace{0.5cm}
    \caption{Constraints satisfied by $\pi$.}
  \label{tab:xyzt}
  \end{table}
\end{proof}

Let $(V,{\mathcal C},k)$ be an instance of {\sc Linear Ordering-AA}, and let $\phi$ be a function from $V$ to $\{0,1,2,3\}$.
For a random $\phi$-compatible linear ordering $\pi$ of $V$, define a binary random variable $z_p$ that takes value one if and only if $C_p$ is satisfied by $\pi$.
Let $Z_p=\mathbb{E}[z_p] -1/6$ for each $p\in [m]$, and let $Z=\sum_{p=1}^mZ_p$.
\begin{lemma}\label{lem:yes}
  If $Z\ge k$ then $(V,{\mathcal C},k)$ is a ``yes''-instance of {\sc Linear Ordering-AA}.
\end{lemma}
\begin{proof}
  By linearity of expectation, $Z\ge k$ implies $\mathbb{E}[\sum_{p=1}^m z_p]\ge m/6 + k$. Thus, if $Z\ge k$ then there is a $\phi$-compatible permutation $\pi$ that satisfies at least $m/6 + k$ constraints.
\end{proof}

Fix a function $\phi:\ V\rightarrow \{0,1,2,3\}$ and assign variables $Y_p,X_p',X_p''$, respectively, to the three instances of {\sc Betweenness} and {\sc Acyclic Subdigraph} above.
\begin{lemma}\label{lem:XYZ}
  For each $p\in [m]$, we have $Z_p=\frac{1}{2}\left[X'_p+X''_p+Y_p\right]$.
\end{lemma}
\begin{proof}
  Let $C_p=(u,v,w)\in {\mathcal C}$.
  Table \ref{tab:XYZ} shows the values of $X'_p,X''_p,Y_p,Z_p$ for some relations between $\phi(u)$, $\phi(v)$ and $\phi(w)$.
  \begin{table}
    \centering
    \begin{tabular}{ccccc}
      \hline
      \noalign{\smallskip}
      Relation between $\phi(u)$, $\phi(v)$ and $\phi(w)$ & $X'_p$ & $X''_p$ & $Y_p$ & $Z_p$\\[0.1cm]
      \noalign{\smallskip}
      \hline
      $\phi(u) = \phi(v)= \phi(w)$                   & 0      & 0       &  0    & 0    \\[0.1cm]
      $\phi(v) < \phi(u)= \phi(w)$                   & -1/2   & 1/2     &  -1/3 & -1/6 \\[0.1cm]
      $\phi(v) > \phi(u)= \phi(w)$                   & 1/2    & -1/2    &  -1/3 & -1/6 \\[0.1cm]
      $\phi(v) = \phi(u) < \phi(w)$                  & 0      & 1/2     &   1/6 & 1/3  \\[0.1cm]
      $\phi(v) = \phi(u) > \phi(w)$                  & 0      & -1/2    &   1/6 & -1/6 \\[0.1cm]
      $\phi(u) < \phi(v)= \phi(w)$                   & 1/2    & 0       &  1/6  & 1/3 \\[0.1cm]
      $\phi(u) > \phi(v)= \phi(w)$                   & -1/2   & 0       &  1/6  & -1/6 \\[0.1cm]
      $\phi(u) < \phi(v) < \phi(w)$                  & 1/2    & 1/2     &  2/3  & 5/6  \\[0.1cm]
      $\phi(u) < \phi(w) < \phi(v)$                  & 1/2    & -1/2    &  -1/3 & -1/6\\[0.1cm]
      $\phi(v) < \phi(u) < \phi(w)$                  & -1/2   & 1/2     &  -1/3 & -1/6 \\[0.1cm]
      $\phi(v) < \phi(w) < \phi(u)$                  & -1/2   & 1/2     &  -1/3 & -1/6 \\[0.1cm]
      $\phi(w) < \phi(u) < \phi(v)$                  &  1/2   & -1/2    &  -1/3 & -1/6 \\[0.1cm]
      $\phi(w) < \phi(v) < \phi(u)$                  & -1/2   & -1/2    &  2/3  & -1/6 \\[0.1cm]
      \noalign{\smallskip}
      \hline
    \end{tabular}
    \vspace{0.2cm}
    \caption{Values of $X'_p,X''_p,Y_p,Z_p$.}
  \label{tab:XYZ}
  \end{table}
  The values of $X'_p,X_p''$ and $Y_p$ can be computed using Tables~\ref{tab:Yp_dist} and~\ref{tab:Xp_dist}.
  In all cases of Table \ref{tab:XYZ} it holds $Z_p=\frac{1}{2}(X'_p+X''_p+Y_p)$.
  Thus, $Z_p=\frac{1}{2}[X'_p+X''_p+Y_p]$ for each possible relation between $\phi(u)$, $\phi(v)$ and $\phi(w)$.
  \end{proof}

Let $X=\sum_{p=1}^m [X'_p+X_p'']$, let $Y=\sum_{p=1}^m Y_p$ and let $\phi$ be a random function from $V$ to $\{0,1,2,3\}$.
Then $X,X'_1,\hdots,X'_m,X''_1,\ldots ,X''_m, Y,Y_1,\hdots,Y_m, Z,Z_1,\ldots ,$ $Z_m$
are random variables.
From $\mathbb{E}[X']=\mathbb{E}[X'']=\mathbb{E}[Y]=0$ it follows that $\mathbb{E}[Z]=0$.

We will be able to use Lemma \ref{lem:polynomial} in the proof of Lemma \ref{lem:yeslem} due to the following:
\begin{lemma}
\label{lem:poly}
  The random variable $Z$ can be expressed as a polynomial of degree 6 in independent  uniformly distributed random variables with values $-1$ and $1.$
\end{lemma}
\begin{proof}
  Consider $C_p = (u,v,w)\in \mathcal C$. Let $\epsilon^u_1=-1$ if $\phi(u)=0$ or 1 and $\epsilon^u_1=1$, otherwise. Let $\epsilon^u_2=-1$ if $\phi(u)=0$ or 2 and $\epsilon^u_2=1$, otherwise. Similarly, we can define $\epsilon^v_1,\epsilon^v_2,\epsilon^w_1,\epsilon^w_2.$ Now $\epsilon^u_1\epsilon^u_2$ can be seen as a binary representation of a number from the set $\{0,1,2,3\}$ and $\epsilon^u_1\epsilon^u_2\epsilon^v_1\epsilon^v_2\epsilon^w_1\epsilon^w_2$ can be viewed as
a binary representation of a number from the set $\{0,1,\ldots ,63\}$, where $-1$ plays the role of 0.
  Then we can write $Z_p$ as the polynomial
  \begin{equation*}
    \frac{1}{64}\sum_{q=0}^{63} (-1)^{s_q}W_q\cdot  (\epsilon^u_1+c^{uq}_1)(\epsilon^{u}_2+c^{uq}_2)(\epsilon^{v}_1+c^{vq}_1)(\epsilon^{v}_2+c^{vq}_2)(\epsilon^{w}_1+c^{wq}_1)(\epsilon^{w}_2+c^{wq}_2),
  \end{equation*}
  where $c^{uq}_1c^{uq}_2c^{vq}_1c^{vq}_2c^{wq}_1c^{wq}_2$ is the binary representation of $q$, $s_q$ is the number of digits equal $-1$ in this representation, and $W_q$ equals the value of $Z_p$ for the case when the binary representations of $\phi(u),\phi(v)$ and $\phi(w)$ are $c^{uq}_1c^{uq}_2$, $c^{vq}_1c^{vq}_2$ and $c^{wq}_1c^{wq}_2$, respectively.
  The actual values for $Z_p$ for each case are given in the proof of Lemma~\ref{lem:XYZ}.
  The above polynomial is of degree 6.
  It remains to recall that $Z=\sum_{p=1}^m Z_p.$

\end{proof}

Let us consider the following natural transformation of our key construction introduced in the beginning of this section. Let $(V,{\mathcal C})$ be an instance of {\sc Linear Ordering} and $(V,\mathcal B)$, $(V,A')$ and $(V,A'')$ be the associated instances of {\sc Betweenness} and {\sc Acyclic Subdigraph}.
Let $b$ be the number of pairs of mutually opposite arcs in the directed multigraph $D=(V,A'\cup A'')$ that are deleted by our reduction rule, and
let $r=2(m-b)$. Let $t$ be the number of complete 3-sets of constraints in $\mathcal B$ whose deletion from $\mathcal B$ eliminates all complete 3-sets of constraints in $\mathcal B$ and let $s=m-3t.$

\begin{lemma}\label{lem:secondmoment}
We have $\mathbb E[Z^2]\ge \frac{11}{3072}(r+s).$
\end{lemma}
\begin{proof}
  Let $A=A'\cup A''=\{a_1,\ldots ,a_{2m}\}$ and $D=(V,A)$.
  Fix a function $\phi:\ V\dom \{0,1,2,3\}$.
  For a random $\phi$-compatible linear ordering $\pi$ of $V$, define a binary random variable $x_i$ that takes value one if and only if $a_i$ is satisfied by $\pi$.
  Analogously, define a binary random variable $y_i$ that takes value one if and only if $B_i$ is satisfied by $\pi$.
  Let $X_i={\mathbb E}[x_i]-1/2$ for all $i = 1,\hdots,2m$, let $Y_j={\mathbb E}[y_j]-1/3$ for all $j=1,\hdots,m$ and let $X=\sum_{i=1}^{2m}X_i$, $Y=\sum_{i=1}^{m}Y_i$.
  Recall that $b$ is the number of deleted pairs of mutually opposite arcs from $D$, and $t$ is the number of complete 3-sets deleted from $\mathcal B$.
  Assume, without loss of generality, that the remaining arcs are $a_1,\hdots,a_r$ and the remaining betweenness constraints are $B_1,\hdots,B_s$.
  Then $X=\sum_{i=1}^{2m} X_i =\sum_{i=1}^rX_i$, $Y=\sum_{i=1}^m Y_i = \sum_{i=1}^s Y_i$ and, by Lemma \ref{lem:XYZ}, $Z=X+Y/2.$ Now let $\phi$ be a random function from $V$ to $\{0,1,2,3\}$.
  We have the following:
    \begin{align*}
  \mathbb E[Z^2] & = \mathbb E[X^2 + XY + Y^2/4]
                 = \mathbb E[X^2] + \mathbb E[Y^2]/4 +  \mathbb E\left[ \left( \sum_{i=1}^r X_i \right) \left( \sum_{j=1}^s Y_j \right) \right]\\
                                  & = \mathbb E[X^2] + \mathbb E[Y^2]/4 + \sum_{i=1}^r \sum_{j=1}^s  \mathbb E[X_iY_j].
  \end{align*}
  We will show that $\mathbb E[X_iY_j]=0$ for any pair $(i,j)$.
  Let $\phi':\ V\dom \{0,1,2,3\}$ be defined as $\phi'(x)=3-\phi(x)$ for all $x$.
  Let $X_i(\phi)$ be the value of $X_i$ when considering $\phi$-compatible orderings and define $X_i(\phi')$, $Y_i(\phi)$ and $Y_i(\phi')$ analogously.
  From Table \ref{tab:Yp_dist} we note that~$Y_j(\phi)=Y_i(\phi')$, and from Table \ref{tab:Xp_dist} we note that~$X_j(\phi)= - X_i(\phi')$.
  From $\mathbb E[X_iY_j] = \frac{1}{4^{|V|}} \sum_{\phi} X_i(\phi) Y_j(\phi)$ it follows that
$$    2\mathbb E[X_iY_j]  = 2 \left[\frac{1}{4^{|V|}} \sum_{\phi} X_i(\phi) Y_j(\phi) \right]
                       = \frac{1}{4^{|V|}} \sum_{\phi} [X_i(\phi) Y_j(\phi) +  X_i(\phi') Y_j(\phi')]=0.$$

  Therefore, $\mathbb E[Z^2] = \mathbb E[X^2] + \mathbb E[Y^2]/4.$
  It follows from Lemmas~\ref{thm:Ysecondmoment} and~\ref{thm:Xsecondmoment} that $\mathbb E[X^2] \ge r/32$ and $\mathbb E[Y^2]\ge \frac{11}{768}s$.
  We conclude that $\mathbb E[Z^2]\ge \frac{11}{3072}(r+s)$.

\end{proof}

\begin{lemma}\label{lem:yeslem}
  There is a constant $c>0$ such that if $r+s\ge ck^2$, then $(V,{\mathcal C},k)$ is a ``yes''-instance of {\sc Linear Ordering-AA}.
\end{lemma}
\begin{proof}
  By Lemmas \ref{lem:poly} and \ref{lem:polynomial}, we have $\mathbb E[Z^4] \leq 9^6(\mathbb E[Z^2])^2$.
  As $\mathbb{E}[Z]=0$, it follows from Lemma~\ref{lem:moments} that $\mathbb P\left(Z > \frac{\sqrt{\mathbb E[Z^2]}}{2\cdot 9^3}\right) > 0$.
  By Lemma \ref{lem:secondmoment}, $\mathbb E[Z^2]\ge \frac{11}{3072}(r+s)$.
  Hence, $\mathbb P\left(Z > \frac{\sqrt{\frac{11}{3072}(r+s)}}{2\cdot 9^3}\right) > 0$.
  Therefore if $r+s\ge ck^2$, where $c=4\cdot 9^6\cdot 3072/11$, then by Lemma \ref{lem:yes}  $(V,{\mathcal C},k)$ is a ``yes''-instance of {\sc Linear Ordering-AA}.
  \end{proof}

After we have deleted mutually opposite arcs from $D$ and complete 3-sets of constraints from  $\mathcal B$ we may assume, by Lemma \ref{lem:yeslem}, that $D$ has an arc multiset $A=\{a_1,\hdots ,a_r\}$ left, with $r= O(k^2)$, and $\mathcal B$ now contains $s=O(k^2)$ constraints $B_1,\hdots ,B_s$.
By Lemma \ref{lem:dev}, ${\rm dev}(V,{\mathcal C})=\max_{\pi}[({\rm dev}(V,A,\pi)+{\rm dev}(V,B,\pi))/2]$, where the maximum is taken over all linear orderings $\pi$ of $V$.

We now create a new instance $(V',\mathcal C',k)$ of {\sc Linear Ordering-AA} as follows.
Let $\omega$ be a new variable not in $V$.
For every $a_i=(u_i,v_i)$ add the constraints $(\omega ,u_i,v_i)$, $(u_i, \omega ,v_i)$ and $(u_i,v_i, \omega )$ to $\mathcal C'$.
For every $B_i=(a_i,\{b_i,c_i\})$ add the constraints $(b_i,a_i,c_i)$ and $(c_i,a_i,b_i)$ to $\mathcal C'$. Let $V'$ be the set of variables that appear in some constraint in $\mathcal C'$.
Then $(V',C')$ is an instance of {\sc Linear Ordering} with $O(k^2)$ variables and constraints.
Now the number of constraints in $\mathcal C'$ satisfied by any linear ordering $\alpha$ of $V'$ equals the number of arcs in $D$ satisfied by $\alpha$ plus the number of constraints in $\mathcal B$ satisfied by $\alpha$.
As the average number of constraints satisfied in $(V',\mathcal C')$ equals $(3r+2s)/6 = r/2 + s/3$, it follows that $\mathsf{dev}(V,{\mathcal C})=\max_{\pi}[(\mathsf{dev}(V,A,\pi)+\mathsf{dev}(V,\mathcal B,\pi))/2] = \mathsf{dev}(V',C')/2$.
Hence, $(V',C',k)$ is a kernel of {\sc Linear Ordering-AA} with $O(k^2)$ variables and constraints.
We have established the following theorem.
\begin{theorem}
\label{LO_AA}
  {\sc Linear Ordering-AA} has a kernel with $O(k^2)$ variables and constraints.
\end{theorem}

Using Proposition \ref{thm:allreducetoone} and Theorem \ref{LO_AA}  we can prove the following:

\begin{theorem}\label{thm:manyi}
There is a bikernel with $O(k^2)$ variables from {\sc $\Pi_i$-AA} to {\sc $\Pi_j$-AA} for each pair $(i,j)$ such that $0\le i\le 10$ and $0\le j\le 10$ but $j\not\in \{2,7\}.$
\end{theorem}
\begin{proof}
By Proposition \ref{thm:allreducetoone}, it suffices to prove this theorem for $i=0$ and $0\le j\le 10$ but $j\not\in \{2,7\}.$ The case $j=0$ follows from Theorem \ref{LO_AA}. Let us consider the remaining cases.\\

{\bf Part 1:}  $j=5$. From the proof of Theorem \ref{LO_AA}, we know that any instance $(V,\mathcal C,k)$ of {\sc Linear Ordering-AA} can be reduced, in polynomial time, to a mixed instance consisting of an instance  $D=(V,A)$ ($|A|=r=O(k^2)$) of {\sc Acyclic Subdigraph} and an instance $(V,\mathcal B)$  ($|\mathcal B|=s=O(k^2)$) of {\sc Betweenness} such that the answer to $(V,\mathcal C,k)$ is ``yes'' if and only if there is a linear ordering of $V$ satisfying, in total, at least $r/2+s/3+k$ arcs and constraints of the mixed instance.  Let $V^*$ be the set of all variables and vertices in constraints of $\mathcal B$ and arcs of $A$. Observe that $|V^*|=O(k^2)$.

Construct an instance $(V',{\mathcal B}',k')$ of {\sc Betweenness-AA} as follows. Set $V'=V^*\cup \{y,z\}$ and initialize ${\mathcal B}'$ by setting ${\mathcal B}'=\mathcal B.$ Add to ${\mathcal B}'$
$(r+s+1)$ copies of the constraint $(x,\{y,z\})$ for each $x\in V^*$ and one copy of the constraint $(v,\{u,z\})$ for each arc $(u,v)\in A$. Observe that $|V'|=O(k^2)$. The total number of constraints in the multiset ${\mathcal B}'$ is $p=(|V^*|+1)(r+s+1)-1$ and recall that the average number of constraints satisfied in an instance of {\sc Betweenness} with $p$ constrains is $p/3.$
We may assume that $p$ is divisible by 3 as otherwise we can add one or two more constraints of the type $(x,\{y,z\})$ to ${\mathcal B}'$.
Let $d=(r+s)-\lceil r/2+s/3+k\rceil$ and let $k'= \frac{2p}{3}-d$. Observe that the answer to $(V',{\mathcal B}',k')$ is ``yes'' if and only if there is a linear ordering of $V'$ that falsifies at most $d$ constraints of ${\mathcal B}'$. Since $d\le r+s$, to falsify at most $d$ constraints of ${\mathcal B}'$, a linear ordering $\alpha$ of $V'$ must satisfy all constraints of the form $(x,\{y,z\})$ and at least $r/2+s/3+k$ other constraints. Since $\alpha$ must satisfy all constraints of the form $(x,\{y,z\})$, we have $\{\alpha^{-1}(1),\alpha^{-1}(|V'|)\}=\{y,z\}$.
Without loss of generality, we may assume that $\alpha^{-1}(|V'|)=z$. Then $\alpha$ satisfies at least $r/2+s/3+k$ other constraints if and only if it satisfies at least $r/2+s/3+k$ arcs and constraints of the mixed instance. Thus, $(V',{\mathcal B}',k')$ is equivalent to $(V,\mathcal C,k)$, and since $k'$ is bounded by a function of $k$, we are done. \\

{\bf Part 2:} $j=1$. Denote constraints of {\sc $\Pi_1$-AA} by $(u<\min\{v,w\})$.
Such a constraint is satisfied by a linear ordering $\alpha$ of $\{u,v,w\}$ if and only if $\alpha(u)<\min\{\alpha(v),\alpha(w)\}$.
Consider the instance $(V',{\mathcal B}',k')$ built in Part 1.  Construct an instance $(V'',{\mathcal C}_1,k_1)$ of {\sc $\Pi_1$-AA} as follows. Let $V''=V'\cup \{z'\}$, where $z'\not\in V'.$
For each constraint $(v,\{u,w\})$ of ${\mathcal B}'$, let ${\mathcal C}_1$ have two copies of $(u<\min\{v,w\})$, two copies of $(w<\min\{u,v\})$ and one copy of $(v<\min\{w,z'\})$ and one copy of $(v<\min\{u,z'\})$.
Thus, ${\mathcal C}_1$ has $6p$ constraints and note that the average number of constraints satisfied in an instance of {\sc $\Pi_1$-AA} with $6p$ constraints is $2p$. Let $k_1=p-d$, where $p$ and $d$ are defined in Part 1.

Let $\alpha$ be a linear ordering of $V''$ and assume that $\alpha$ satisfies the maximum number of constraints in ${\mathcal C}_1$ and this number is at least $2p+k_1=3p-d$. We may assume that $\alpha(z')=|V''|$ as moving $z'$ to the last position in the linear ordering will not falsify any constraint of ${\mathcal C}_1$.
Observe now that if $\alpha$ satisfies $(v,\{u,w\})$, then it satisfies exactly three constraints of ${\mathcal C}_1$
from the six constraints generated by $(v,\{u,w\})$ and if $\alpha$ falsifies $(v,\{u,w\})$, it satisfies exactly two constraints of ${\mathcal C}_1$
from the six constraints generated by $(v,\{u,w\})$. Therefore, $\alpha$ satisfies exactly $3t+2(p-t)$ constraints of ${\mathcal C}_1$, where $t$ is the number of constraints in ${\mathcal B}'$ satisfied by $\alpha$. Hence, $t\ge p-d$.

Now assume that a linear ordering $\alpha$ of $V'$ satisfies at least $p-d$ constraints of ${\mathcal B}'$. We extend $\alpha$ to $V''$ by setting $\alpha(z')=|V''|$. Similarly to the above we can show that $\alpha$ satisfies at least $2p+k_1=3p-d$ constraints in ${\mathcal C}_1$.
Thus, $(V',{\mathcal C}_1,k_1)$ is equivalent to $(V',{\mathcal B}',k')$ and, therefore by Part 1, to $(V,\mathcal C,k)$, an instance of {\sc Linear Ordering-AA}. Clearly, $|V''|=O(k^2)$ and $k_1$ is bounded by a function of $k$. \\

{\bf Part 3:} $j=3$. In Part 2, we have proved that for any instance $(V,\mathcal C,k)$ of {\sc Linear Ordering-AA} there is an equivalent instance $(V',{\mathcal C}_1,k_1)$ of $\Pi_1$-AA
with $O(k^2)$ variables and distinct constraints (and $k_1$ is bounded by a function of $k$). Recall that $(V',{\mathcal C}_1,k_1)$ has $6p$ constraints. Let $\alpha$ be a linear ordering
of $V'$ and let $\alpha'$ be the reverse ordering. As in the proof of Case $i=3$ of Theorem \ref{thm:allNPhard}, construct from $(V',{\mathcal C}_1,k_1)$
an instance $(V',{\mathcal C}_3,k_3)$ of {\sc $\Pi_3$-AA} such that ${\mathcal C}_3$ has $12p$ constraints and
at least $q$ constraints of $\mathcal C_1$ are satisfied by $\alpha$
if and only if at least $2q+(|\mathcal C_1|-q)$ constraints of $\mathcal C_3$ are satisfied in $\alpha'$. Let $q=2p+k_1$ and $k_3=k_1$. Assume that $(V',{\mathcal C}_1,k_1)$ is a ``yes''-instance certified by $\alpha$. Then $\alpha'$ satisfies at least $8p+k_3$ constraints of $(V',{\mathcal C}_3,k_3)$ and $(V',{\mathcal C}_3,k_3)$ is a ``yes''-instance. Similarly, if $(V',{\mathcal C}_3,k_3)$ is a ``yes''-instance, then $(V',{\mathcal C}_1,k_1)$ is a ``yes''-instance, too. \\

{\bf Part 4:} $j=4,8,9,10$. For each $j=4,8,9,10$ the proof is similar to Part 2 and, thus, we will only describe how to transform the instance $(V',{\mathcal B}',k')$ built in Part 1 into an instance $(V',{\mathcal C}_i,k')$ of {\sc $\Pi_i$-AA} for every $i=4,8,9,10$,  and observe how the fact that a constraint $B$ of $(V',{\mathcal B}',k')$ is satisfied or falsified corresponds to the number of satisfied constraints in the instance of {\sc $\Pi_i$-AA} generated by $B.$ Then it is not hard to check that $(V',{\mathcal B}',k')$ and $(V',{\mathcal C}_i,k')$ are equivalent.\\

\noindent{\bf Case $j=4$}. Denote constraints of {\sc $\Pi_4$-AA} by $(u\ \| \{v<w\}))$.
Such a constraint is $\Pi_4$-satisfied by a linear ordering $\alpha$ of $\{u,v,w\}$ if and only if $\alpha(v)<\alpha(w)$ and $\alpha(u)$ is not between $\alpha(v)$ and $\alpha(w)$. Construct an instance $(V',{\mathcal C}_4,k_4)$ of {\sc $\Pi_4$-AA} as follows. For each constraint $(v,\{u,w\})$ of ${\mathcal B}'$, let ${\mathcal C}_4$ have four constraints: $(u\ \| \{v<w\}))$, $(u\ \| \{w<v\}))$, $(w\ \| \{u<v\})$ and $(w\ \| \{v<u\})$.  It is easy to check that if $(v,\{u,w\})$ is satisfied by a linear ordering $\alpha$ of $V'$, then two of the four constraints are satisfied by $\alpha$ and if $(v,\{u,w\})$ is falsified by $\alpha$, then only one of the four constraints is satisfied by $\alpha$.\\

\noindent{\bf Case $j=8$}. Denote constraints of {\sc $\Pi_8$-AA} by $(v<u<w \mbox{ or } w<v)$. Such a constraint is satisfied by a linear ordering $\alpha$ of $\{u,v,w\}$ if and only if either $\alpha(v)<\alpha(u)<\alpha(w)$ or $\alpha(w)<\alpha(v)$. For each constraint $(v,\{u,w\})$ of ${\mathcal B}'$, let ${\mathcal C}_8$ have two constraints: $(w<v<u \mbox{ or } u<w)$ and $(u<v<w \mbox{ or } w<u).$  It is easy to check that if $(v,\{u,w\})$ is satisfied by a linear ordering $\alpha$ of $V'$, then both constraints generated by $(v,\{u,w\})$ are satisfied by $\alpha$ and if $(v,\{u,w\})$ is falsified by $\alpha$, then only one of two constraints is satisfied by $\alpha$.\\

\noindent{\bf Case $j=9$}. Denote constraints of {\sc $\Pi_9$-AA} by $(v\ \| \{u,w\}))$. Such a constraint is satisfied by a linear ordering $\alpha$ of $\{u,v,w\}$ if and only if $\alpha(v)$ is not between $\alpha(u)$ and $\alpha(w)$. Construct an instance $(V',{\mathcal C}_9,k_9)$ of {\sc $\Pi_9$-AA} as follows. For each constraint $(v,\{u,w\})$ of ${\mathcal B}'$, let ${\mathcal C}_9$ have two constraints: $(u\ \| \{v,w\}))$ and $(w\ \| \{u,v\}))$.  It is easy to check that if $(v,\{u,w\})$ is satisfied by a linear ordering $\alpha$ of $V'$, then both constraints generated by $(v,\{u,w\})$ are satisfied by $\alpha$ and if $(v,\{u,w\})$ is falsified by $\alpha$, then only one of two constraints is satisfied by $\alpha$.\\

\noindent{\bf Case $j=10$}. Denote constraints of {\sc $\Pi_{10}$-AA} by $(\mbox{not } u<v<w)$. Such a constraint is satisfied by a linear ordering $\alpha$ of $\{u,v,w\}$ if and only if
we do not have $\alpha(u)<\alpha(v)<\alpha(w)$.  For each constraint $(v,\{u,w\})$ of ${\mathcal B}'$, let ${\mathcal C}_{10}$ have four constraints: $(\mbox{not } v<u<w)$, $(\mbox{not } v<w<u)$, $(\mbox{not } u<w<v)$ and $(\mbox{not } w<u<v)$. It is easy to check that if $(v,\{u,w\})$ is satisfied by a linear ordering $\alpha$ of $V'$, then all four constraints generated by $(v,\{u,w\})$ are satisfied by $\alpha$ and if $(v,\{u,w\})$ is falsified by $\alpha$, then only three of the four constraints are satisfied by $\alpha$.

\vspace{3mm}

{\bf Part 5:} $j=6$. Denote constraints of {\sc $\Pi_6$-AA} by $(u<v<w \mbox{ or } w, \{u,v\})$. Such a constraint is satisfied by a linear ordering $\alpha$ of $\{u,v,w\}$ if and only if either $\alpha(u)<\alpha(v)<\alpha(w)$ or $\alpha(w)$ is between $\alpha(u)$ and $\alpha(v)$. Consider the instance $(V',{\mathcal B}',k')$ built in Part 1. Construct an instance $(V_6,{\mathcal C}_6,k_6)$ of {\sc $\Pi_6$-AA} as follows.

Let $V_6=V'\cup \{a,b\}$, where $\{a,b\}\cap V'=\emptyset$. Initiate ${\mathcal C}_6$ by adding to it, for each $x\in V'$, $6p+1$ copies of $(x<b<a \mbox{ or } a,\{x,b\})$ and $6p+1$ copies of $(x<a<b \mbox{ or } b,\{x,a\})$. For each $(v,\{u,w\})\in {\mathcal B}'$, add to ${\mathcal C}_6$ the following constraints: two copies of $(u<w<v \mbox{ or } v,\{u,w\})$, two copies of $(w<u<v \mbox{ or } v,\{u,w\})$, a copy of $(b<v<u \mbox{ or } u,\{v,b\})$, and a copy of $(b<v<w \mbox{ or } w,\{b,v\})$. Recall that ${\mathcal B}'$ has $p$ constraints and note that ${\mathcal C}_6$ has $6p + 2(6p+1)|V'|$ constraints. Observe that the average number of satisfied constraints, in an instance of {\sc $\Pi_6$-AA} with $6p + 2(6p+1)|V'|$ constraints, is $3p + (6p+1)|V'|$. Let $k_6=(6p+1)|V'|+(2p-3d)$, where $d$ is defined in Part 1.

Then $(V_6,{\mathcal C}_6,k_6)$ is a ``yes''-instance if and only if there is a linear ordering $\alpha$ of $V_6$ that satisfies at least $2(6p+1)|V'|+(5p-3d)$ constraints.
For $\alpha$ to satisfy so many constraints, it must satisfy all constraints of the forms $(x<b<a \mbox{ or } a,\{x,b\})$ and  $(x<a<b \mbox{ or } b,\{x,a\})$, implying that $a$ and $b$ must be the last two variables in $\alpha$, and at least $5p-3d$ constraints generated by ${\mathcal B}'$. Observe that if $\alpha$ satisfies $(v,\{u,w\})\in {\mathcal B}'$ then exactly five constraints of ${\mathcal C}_6$ generated by $(v,\{u,w\})$ are satisfied by $\alpha$ and if $\alpha$ falsifies $(v,\{u,w\})\in {\mathcal B}'$ then exactly two constraints of ${\mathcal C}_6$ generated by $(v,\{u,w\})$ are satisfied by $\alpha$. Thus, $\alpha$ satisfies at least $5p-3d$ constraints generated by ${\mathcal B}'$ if and only if $\alpha$ satisfies at least $p-d$ constraints of ${\mathcal B}'$. Therefore, $(V',{\mathcal B}',k')$ and $(V_6,{\mathcal C}_6,k_6)$ are equivalent.
\end{proof}

Using Theorems \ref{thm:ASDkernel} and \ref{thm:manyi} we can prove the following:

\begin{theorem}\label{thm:main}
All ternary Permutation-CSPs parameterized above average have  kernels with $O(k^2)$ variables.
\end{theorem}
\begin{proof}
By Theorem \ref{thm:manyi}, it suffices to prove that the problems {\sc $\Pi_j$-AA}, $j=2,7$,  have kernels with quadratic number of variables.

\noindent{\bf Case $j=2$}. Denote constraints of {\sc $\Pi_2$-AA} by $(u,v<w)$.
Such a constraint is satisfied by a linear ordering $\alpha$ of $\{u,v,w\}$ if and only if $\alpha(v) < \alpha(w)$.
Consider the instance $(V,{\mathcal C},k)$ of {\sc $\Pi_2$-AA} and construct an instance $(V,A,k)$ of {\sc Acyclic Subdigraph-AA} as follows:
if  $(u,v<w)\in \mathcal C$ then $(v,w)$ is added to $A.$ Clearly, $(V,{\mathcal C},k)$ and $(V,A,k)$ are equivalent. By Theorem \ref{thm:ASDkernel}, in polynomial time, $(V,A,k)$ can be transformed into an equivalent instance $(V',A',k')$ of {\sc Acyclic Subdigraph-AA} such that $|V'|=O(k^2)$ and $k'$ is bounded by a function of $k$ (in fact, $k'=k$). As in the proof of Case $i=2$ of Theorem \ref{thm:allNPhard}, from $(V',A',k')$ we can construct an equivalent instance $(V^*,{\mathcal C}^*,k')$ of {\sc $\Pi_2$-AA} such that $|V^*|=|V'|+1=O(k^2)$. Observe that $(V^*,{\mathcal C}^*,k')$ is the required kernel.\\

\noindent{\bf Case $j=7$}. Denote constraints of {\sc $\Pi_7$-AA} by $\langle u,v,w \rangle$. Such a constraint is satisfied by a linear ordering $\alpha$ of $\{u,v,w\}$ if and only if either $\alpha(u)<\alpha(v) < \alpha(w)$ or $\alpha(v)<\alpha(w) < \alpha(u)$ or $\alpha(w)<\alpha(u) < \alpha(v)$. Consider the instance $(V,{\mathcal C},k)$ of {\sc $\Pi_7$-AA} and construct an instance $(V,A,k)$ of {\sc Acyclic Subdigraph-AA} as follows: if  $\langle u,v,w \rangle \in \mathcal C$ then $(u,v),(v,w)$ and $(w,u)$ are added to $A.$ Let $\alpha$ be a linear ordering of $V$ and observe that if $\langle u,v,w \rangle$ is satisfied by $\alpha$ then exactly two of the three arcs of $A$ generated by $\langle u,v,w \rangle$ are satisfied by $\alpha$ and if $\langle u,v,w \rangle$ is falsified by $\alpha$ then exactly one of the three arcs of $A$ generated by $\langle u,v,w \rangle$ is satisfied by $\alpha$. Thus, $\alpha$ satisfies at least $|\mathcal C|/2 + k$ constraints of ${\mathcal C}$ if and only if $\alpha$ satisfies at least $2(|\mathcal C|/2+k)+ (|\mathcal C|/2-k) = 3|\mathcal C|/2 + k = |A|/2 + k$ arcs of $A$.
By Theorem \ref{thm:ASDkernel}, in polynomial time, $(V,A,k)$ can be transformed into an equivalent instance $(V',A',k')$ of {\sc Acyclic Subdigraph-AA} such that $|V'|=O(k^2)$ and $k'$ is bounded by a function of $k$ (in fact, $k'=k$).

Now construct an instance $(V'',{\mathcal C}',k')$ of {\sc $\Pi_7$-AA} by setting $V''=V'\cup \{z\}$, where $z\not\in V'$, and ${\mathcal C}'=\{\langle u,v,z \rangle :\ (u,v)\in A'\}$.
Let $\alpha$ be a linear ordering of $V''$ satisfying at least $|{\mathcal C}'|/2+k'$ constraints of ${\mathcal C}'.$  We may assume that $\alpha(z)=|V''|$ as moving the last element of an ordering to the front of the ordering does not falsify any constraint, and so by repeatedly doing this we will move $z'$ to the last position in our ordering.
Thus, $\alpha$ satisfies at least  $|A'|/2+k'$ arcs of $A'$. Now let $\alpha$ be a linear ordering of $V'$ satisfying at least $|A'|/2+k'$ arcs of $A'$. Extend $\alpha$ to $V''$ by setting $\alpha(z)=|V''|$ and observe that $\alpha$ satisfies at least $|{\mathcal C}'|/2+k'$ constraints in
${\mathcal C}'$. Hence, $(V'',{\mathcal C}',k')$ is equivalent to $(V',A',k')$ and, thus, to $(V,{\mathcal C},k)$ implying that $(V'',{\mathcal C}',k')$ is a kernel of {\sc $\Pi_7$-AA}.
\end{proof}

\section{Normal Reduction Rules for Linear Ordering-AA}\label{sec:inf}

We call a reduction rule {\em normal} if it removes a
number of constraints which will always have the average number of constraints satisfied
no matter what ordering is used.
Note that all reduction rules for
{\sc Betweenness-AA} and {\sc Acyclic Subdigraph-AA} are normal.

Theorem \ref{lem2} below implies that infinite number of instances of {\sc Linear Ordering} cannot be reduced by any normal reduction rule, except the one
that removes all constraints in the instance.  Therefore, no finite set of normal reduction rules can guarantee that one always gets
either the empty instance or an instance where one can do better than the average.
For both {\sc Betweenness-AA} and {\sc Acyclic Subdigraph-AA} we only needed one normal reduction rule to get such a guarantee. This is another
indication that {\sc Linear Ordering-AA} is a more difficult problem.

Let us describe a directed graph $G_i$ with vertex set $V_i$ and a decomposition, $C_i$,  of the arc set of $G_i$ into directed $3$-cycles.
When $i=0$ we have $V_0=\{x_1,x_2,x_3\}$  and $C_0=\{ x_1x_2x_3x_1, x_3x_2x_1x_3 \}$. Note that the arc set of $G_i$ is always the set of arcs used in
$C_i$.

 When $i>0$ we will construct $G_i$, $V_i$ and $C_i$ recursively.
 So assume that $G_{i-1}$, $V_{i-1}$ and $C_{i-1}$ have been constructed and let $G_{i-1}'$
be another copy of $G_{i-1}$ on vertex set $V_{i-1}'$ and with decomposition $C_{i-1}'$.
Let $V_i = V_{i-1} \cup V_{i-1}'$ and note that $|V_i|=2|V_{i-1}|$.
Let $c=x_a x_b x_c x_a$ be any directed $3$-cycle in $C_{i-1}$ and let $c'=x_d' x_e' x_f' x_d'$ be any directed
$3$-cycle in $C_{i-1}'$. Let $C_i$ contain all directed $3$-cycles in $C_{i-1}\setminus \{c\}$ and $C_{i-1}'\setminus \{c'\}$ and
the following six directed $3$-cycles:

\begin{center}
$\begin{array}{ccccc}
c_1 =  x_a  x_b  x_f' x_a, &  \hspace{0.5cm} &
c_2 =  x_b  x_c  x_e' x_b, &   \hspace{0.5cm} &
c_3 =  x_c  x_a  x_d' x_c,  \\
c_4 =  x_d' x_e' x_c  x_d', &  &
c_5 =  x_e' x_f' x_b  x_e', &  &
c_6 =  x_f' x_d' x_a  x_f'.\\
\end{array} $
\end{center}

A directed graph $D=(V,A)$ is {\em symmetric} if $(u,v)\in A$ implies $(v,u)\in A.$

\begin{lemma} \label{lem1}
We have that   $|V_i|=3 \times 2^i$ and that $G_i$ is a symmetric digraph with no parallel arcs for all $i \geq 0$.
  Furthermore if $C^*_i$ is a proper nonempty subset of $C_i$ then the arcs of $C^*_i$ do not form a symmetric digraph.
\end{lemma}

\begin{proof}
Since $|V_0|=3$ and $|V_i|=2  |V_{i-1}|$ we have $|V_i|=3 \times 2^i$ for all $i \geq 0$.
Clearly $G_0$ is symmetric  with no parallel arcs. Assume that $G_j$ is symmetric with no parallel arcs for each $0\le j<i$ and consider $G_i$, $i>0$.
It is not difficult to see that by deleting the arcs in $c$ and $c'$ and adding the
arcs in $c_1,c_2, \ldots,c_6$ we obtain a symmetric digraph with no parallel arcs, which completes the proof of
the first part of the lemma.

The second part of the lemma clearly holds when $i=0$, so assume that $i>0$ and that the second part holds for each $0\le j<i$.
If $C^*_i \cap \{c_1,c_2,c_3,c_4,c_5,c_6\} = \emptyset$ then we are done by induction
as either $C^*_i \cap C_{i-1}$ or $C^*_i \cap C_{i-1}'$ is non-empty and therefore induces a non-symmetric subdigraph.

So we may assume that $C^*_i \cap \{c_1,c_2,c_3,c_4,c_5,c_6\} \not= \emptyset$. Suppose that the arcs of of $C^*_i$ form a
symmetric digraph.
Due to the connection between $x_a$ and $x_f'$ we note that $c_1 \in C_i^*$ if and only if $c_6 \in C_i^*$.
Analogously, $c_1 \in C_i^*$ if and only if $c_5 \in C_i^*$ (due to $x_f' x_b$), $c_2 \in C_i^*$ if and
only if $c_4 \in C_i^*$ (due to $x_e' x_c$), $c_2 \in C_i^*$ if and
only if $c_5 \in C_i^*$ (due to $x_e' x_b$), $c_3 \in C_i^*$ if and
only if $c_6 \in C_i^*$ (due to $x_d' x_a$), and $c_3 \in C_i^*$ if and
only if $c_4 \in C_i^*$ (due to $x_d' x_c$).
Thus, if $C^*_i \cap \{c_1,c_2,c_3,c_4,c_5,c_6\} \not= \emptyset$ and the arcs of $C^*_i$ form a
symmetric digraph then we must always have  $c_1,c_2,c_3,c_4,c_5,c_6
\in C_i^*$.

As $C^*_i$ is a proper subset of $C_i$ we may without loss of generality assume that
there is a directed $3$-cycle in $C_{i-1}\setminus \{c\}$ (otherwise it is in $C_{i-1}'\setminus \{c'\}$)
which does not belong to $C_i^*$ and by induction the arc set of
$(\{c\} \cup C^*_i) \cap C_{i-1}$ does not form a symmetric digraph. Therefore the arcs of $C^*_i$ do not form a symmetric digraph either, a contradiction.
This completes the proof of the lemma.
\end{proof}

For each $i \geq 0$ we construct an instance $(V_i,K_i)$ of {\sc Linear Ordering-AA} as follows.
For every directed $3$-cycle in $C_i$, say $u v w u$, add the following three constraints
$(u,v,w)$, $(v,w,u)$ and $(w,u,v)$ to $K_i$.
Let $(V_i,\mathcal B_i)$ be the instance of {\sc Betweenness-AA} which we associate with $(V_i,K_i)$ in Section  \ref{sec:LAAA}
and let $(V_i,A_i')$ and $(V_i,A_i'')$ be the two instances of {\sc Acyclic Subdigraph-AA} which we also associate with $(V_i,K_i)$ there.
By Lemma \ref{lem:dev}, the following holds for all linear orderings $\alpha$ of $V_i$:

  \begin{equation}\label{Lem7eq}
    \mathsf{dev}(V_i,K_i,\alpha)
      = \frac{1}{2}\left[   \mathsf{dev}(V_i,A_i',\alpha)
                          + \mathsf{dev}(V_i,A_i'',\alpha)
                          + \mathsf{dev}(V_i,{\mathcal B_i},\alpha)
                  \right].
  \end{equation}

\begin{theorem}\label{lem2}
We have  $\mathsf{dev}(V_i,K_i) = 0$ and if  $K_i^*$ is a nonempty proper subset of $K_i$ then we can always satisfy more than $|K_i^*|/6$
constraints of $K_i^*$.
\end{theorem}
\begin{proof}
 As a directed $3$-cycle $u v w u$ in $C_i$ gives rise to the betweenness constraints $(v,\{u,w\})$, $(w,\{v,u\})$ and $(u,\{w,v\})$ in $\mathcal B_i$
we can only satisfy $|C_i|$ constraints in $\mathcal B_i$. Furthermore, a directed $3$-cycle $u v w u$ in $C_i$ gives rise to two copies of the
constraints $(u,v)$, $(v,w)$ and $(w,u)$ in $A_i' \cup A_i''$. Thus, we can think of an arc, $uv$,  in $G_i$ as giving rise to two copies of the
acyclic subdigraph constraint $(u,v)$. As $G_i$ is symmetric this means that every constraint $(u,v)$ can be paired with  a constraint $(v,u)$ so
we can only satisfy half the constraints in $A'_i \cup A_i''$. As we can only satisfy the average number of constraints in both $A_i' \cup A_i''$ and $B_i$,
(\ref{Lem7eq}) implies that $\mathsf{dev}(V_i,K_i)=0$, which proves the first part of the lemma.

For the sake of contradiction assume that  $K_i^*$ is a nonempty proper subset of $K_i$ and
that $\mathsf{dev}(V_i,K_i^*) = 0$.
Let $(V_i,\mathcal B_i^*)$ be the instance of {\sc Betweenness-AA} which we associate with $(V_i,K_i^*)$ in Section  \ref{sec:LAAA}
and let $(V_i,A_i^*)$ and $(V_i,A_i^{**})$ be the two instances of {\sc Acyclic Subdigraph-AA} which are also associated with $(V_i,K_i^*)$.
Let $Z$, $Y$ and $X$ be the random variables associated with $(V_i,K_i^*)$, $(V_i,\mathcal B_i^*)$ and $(V_i,A_i^* \cup A_i^{**})$, respectively.
Note that $\mathsf{dev}(V_i,K_i^*) = 0$ is equivalent to $\mathbb E[Z^2]=0$, which by the proof of
Lemma \ref{lem:secondmoment} implies
that $\mathbb E[X^2]=0$ and $\mathbb E[Y^2]=0$.
Observe that by Lemma \ref{thm:Ysecondmoment}  this implies that if $(u,\{v,w\}) \in \mathcal B_i^*$ then
$(w,\{v,u\}), (v,\{u,w\}) \in \mathcal B_i^*$.
So, if $(u,v,w) \in K_i^*$, then  $(v,w,u), (w,u,v) \in K_i^*$.
Therefore, $K_i^*$ can be thought of as being obtained from a proper subset, $C_i^*$, of
the directed $3$-cycles $C_i$.  Observe that by Lemma \ref{lem1} some arc $(u,v)$ belongs to a directed $3$-cycle in $C_i^*$, but the arc $vu$ does not belong
to such a directed $3$-cycle. However, this implies that $(u,v) \in A_i^* \cup A_i^{**}$, but $(v,u) \not\in A_i^* \cup A_i^{**}$. Thus,
$\mathbb E[X^2]>0$ by Lemma \ref{thm:Xsecondmoment}. This contradiction completes the proof.
\end{proof}

\section{Further Research}\label{sec:fr}

It is natural and easy to extend the definition of $\Pi$-AA to a fixed arity $r> 3$.
Similar to Proposition \ref{thm:allreducetoone}, it is easy to prove
that, for each fixed $r$ every $\Pi$-AA can be reduced to $\Pi_0$-AA, where $\Pi_0$ is
{\sc Linear Ordering} of arity $r$. However, it appears technically very difficult to extend results obtained for arities $r=2$ and 3 to $r>3$. We conjecture that for each fixed $r$ all problems $\Pi$-AA are fixed-parameter tractable.

We have parameterized {\sc Linear Ordering} of any arity $r$ using the average as a tight lower bound. Similarly, we can parameterize {\sc Linear Ordering} below a tight upper bound and the number of constraints $m$ can be set as a tight upper bound. So, the problem is whether there is a bijection $\alpha:\ V\rightarrow [n]$ which satisfies at least $m-k$ constraints of an instance $(V,\mathcal C)$ of {\sc Linear Ordering}, where $k$ is the parameter.
It is easy to show that for $k=0$ the problem is polynomial-time solvable, but
it seems to be a difficult question to determine the parameterized complexity of this problem for any arity $r\ge 2$.

Note that for arity $r=2$ the corresponding problem is {\sc Directed Feedback Arc Set} parameterized below the number $m$ of arcs in a given directed graph. The parameterized complexity of the last problem was an open question for many years \cite{GutinYeoSurvey} and, only in 2008, Chen et al. \cite{Chen} proved that the problem is fixed-parameter tractable. (It is still unknown whether the last problems admits a polynomial-size kernel.) For every fixed arity $r\ge 3$, the parameterized complexity of {\sc Linear Ordering} parameterized below $m$ is unknown.

\end{document}